\documentstyle[12pt]{article}
\title{Do $H_0$ and $q_0$ really have the values we believe they
have?}
\author{Ll. Bel \\
Lab. Gravitation et Cosmology Relativistes. ESA 70 \\
{\small Tour 22-12, 4 place Jussieu, 75252 Paris}
}
\date{\today}
\begin{document}
\maketitle

\begin{abstract}
We present an example where a justified modification of the law of
propagation of light in a Robertson-Walker model of the universe
leads to an identification of $H_0$ and $q_0$ different from that
corresponding to the usual law of propagation along null geodesics.
We conclude from this example that observed values which we would  
associate with the values of $H_0$ and $q_0$ with the usual
interpretation correspond in fact to the values of $2H_0$ and
$\frac12(q_0-1)$. It is therefore possible that observed values that
we usually interpret as corresponding to a moderately aged 
universe with accelerating expansion may in fact correspond
a much older universe with a decelerating expansion.
\end{abstract}
 
When astronomers say that they obtain from observations the values of
the Hubble constant and deceleration parameter of the universe they
mean that they measure the slope $H^*_0$ and a curvature related
parameter $q^*_0$ at the origin of the Hubble graph:

\begin{equation}
\label {1.1}
z=H^*_0d+\frac12H_0^{*2}(1+q_0)d^2
\end{equation}  
where $z$ is the red-shift and $d$ is an
operationally defined, agreed upon, distance indicator of observed
sources.

To confront the observed Hubble graph with theory requires the
description of a cosmological model and the theoretical
identification of the two parameters $H^*_0$ and $q^*_0$, and the 
distance indicator $d$. The simplest cosmological models assume a
Robertson-Walker model with line-element:

\begin{equation}
\label {1.2}
d\tau^2=-dt^2+F^2(t)N^2(r)\delta_{ij}dx^idx^j, \quad
N(r)=1/(1+kr^2/4), \quad r=\sqrt{\delta_{kl}x^kx^l}
\end{equation}
where $F(t)$ is the scale factor, $k$ the curvature of space, and
where units have been chosen such that the universal speed constant $c$
is equal to $1$.

Assuming also that
light behaves in a Robertson-Walker space-time as it does in a pure
vacuum domain with minimal coupling of electromagnetism to gravitation,
i.e. 
assuming that light propagates along null geodesics of the line-element
above
one derives (\ref{1.1}) with the following identification:

\begin{equation}
\label {1.3}
H^*_0=H_0\equiv \frac{\dot F}{F}, \quad
q^*_0=q_0\equiv -\frac{\ddot FF}{{\dot F}^2}
\end{equation}  
and:

\begin{equation}
\label {1.3a}
d=\int_0^r N(r)\, dr
\end{equation}
 
This point of view has been questioned
recently in two papers (Refs. \cite{Albrecht-Magueijo} and \cite{Barrow})
that consider 
cosmological models
with a varying speed of light, either in a framework more general
than general relativity, like scalar-tensor theories of gravity 
or in more general
phenomenological approaches.  

On the other hand it is also possible to consider a varying speed of
light in the framework of general relativity as it was discussed in
our recent gr-qc preprint (Ref \cite{Froth}). This paper is based on
the idea that $F$ in the line-element (\ref{1.2}) can be interpreted
as the inverse of an effective speed of light or equivalently, as a
refractive index. In this case the theory of light propagation in a
non-dispersive medium (See for instance \cite{Synge}) 
tell us that the light rays are the null geodesics
of the metric:

\begin{equation}
\label {1.4}
\bar g_{\alpha\beta}=g_{\alpha\beta}+(1-F^{-2})u_\alpha u_\beta
\end{equation}     
where $g_{\alpha\beta}$ is the metric corresponding to the
line-element (\ref{1.2}) and $u^\alpha$ is the time-like unit vector
with components $u^0=1$ and $u^i=0$ in the corresponding
coordinates. Therefore:

\begin{equation}
\label {1.4b}
\bar g_{00}=-F^{-2}, \quad \bar g_{0i}=0, \quad \bar
g_{ij}=F^2N^2\delta{ij}
\end{equation}
This means that a light ray emanating
from any point with radial coordinate $r_e$ at time $t_e$
reaches the point with radial coordinate $r=0$ at time $t$ given by:

\begin{equation} 
\label {1.4a} \int_0^{r_e} N(r)\, dr=\int_{t_e}^tF^{-2}(t)\, dt 
\end{equation} 
With the usual convention about light propagation the integrand in the
r-h-s is $F^{-1}$
instead of $F^{-2}$.

The interesting fact that comes out from this interpretation is that
the same elementary calculation that leads to the relationship
(\ref{1.1}) with the identification (\ref{1.3}), when one assumes that
light 
propagates along geodesics of (\ref{1.2}) now leads to the following
identification:

\begin{equation}
\label {1.5}
H^*_0=2H_0, \quad q^*_0=\frac12(q_0-1) 
\end{equation}
in which case the dynamics of the universe becomes radically
different from what it is believed to be now.
Let us assume as a simple numerical example that:  
  
\begin{equation}
\label {A.4}
q^*_0=-0.1, \quad H^*_0=60\, \mbox{km/s/Mpc}
\end{equation}
are the observed values of the parameters in (\ref{1.1}). According to
the usual identification (\ref{1.3}) this corresponds to a moderately
aged universe ($H^{-1}_0=17\times 10^9\,$yr) with accelerating
expansion, while with the new identification (\ref{1.5}) the 
values of $q_0$ and $H_0$ are:

\begin{equation}
\label {A.5}
q_0=+0.8 \quad H_0=30\, \mbox{km/s/Mpc}
\end{equation}
which correspond to a very old universe ($H^{-1}_0=33\times
10^9\,$yr) with a decelerating expansion.

All this should remind us that modern cosmology
is still a very young branch of physics and astronomy where very much
remains to be discovered and clarified.

\end{document}